\documentclass{WileyMSP-template}

\usepackage{authblk}
\usepackage{subfigure}
\usepackage{hyperref}
\usepackage{amsmath}
\usepackage[backend=biber,sorting=none]{biblatex}
\usepackage{verbatim}
\usepackage{placeins}
\usepackage{csquotes}
\usepackage{xcolor}

\pdfoutput=1
\begin{document}

\pagestyle{fancy}
\rhead{\includegraphics[width=2.5cm]{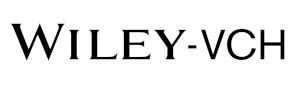}}

\title{Fast shimming algorithm based on Bayesian optimization\\ for magnetic resonance based dark matter search}
\maketitle

\author{Julian Walter}
\author{Hendrik Bekker}
\author{John Blanchard}
\author{Dmitry Budker}
\author{Nataniel L. Figueroa}
\author{Arne Wickenbrock}
\author{Yuzhe Zhang}
\author{Pengyu Zhou}

\begin{affiliations}
Julian Walter\\
Johannes Gutenberg-Universit{\"a}t Mainz, 55128 Mainz, Germany\\
Helmholtz-Institut, GSI Helmholtzzentrum f{\"u}r Schwerionenforschung, 55128 Mainz, Germany\\
Email Address: \textit{juwalter@students.uni-mainz.de}\\
\,\\
Dr. Hendrik Bekker\\
Johannes Gutenberg-Universit{\"a}t Mainz, 55128 Mainz, Germany\\
\,\\
Dr. John Blanchard\\
Quantum Technology Center, University of Maryland, College Park, Maryland 20742, USA\\
\,\\
Prof. Dr. Dmitry Budker\\
Johannes Gutenberg-Universit{\"a}t Mainz, 55128 Mainz, Germany\\
Helmholtz-Institut, GSI Helmholtzzentrum f{\"u}r Schwerionenforschung, 55128 Mainz, Germany\\
Department of Physics, University of California, Berkeley, CA 94720-7300, United States of America\\
\,\\
Dr. Nataniel L. Figueroa\\
Johannes Gutenberg-Universit{\"a}t Mainz, 55128 Mainz, Germany\\
Helmholtz-Institut, GSI Helmholtzzentrum f{\"u}r Schwerionenforschung, 55128 Mainz, Germany\\
\,\\
Dr. Arne Wickenbrock\\
Johannes Gutenberg-Universit{\"a}t Mainz, 55128 Mainz, Germany\\
Helmholtz-Institut, GSI Helmholtzzentrum f{\"u}r Schwerionenforschung, 55128 Mainz, Germany\\
\,\\
Yuzhe Zhang\\
Johannes Gutenberg-Universit{\"a}t Mainz, 55128 Mainz, Germany\\
Helmholtz-Institut, GSI Helmholtzzentrum f{\"u}r Schwerionenforschung, 55128 Mainz, Germany\\
\,\\
Pengyu Zhou\\
Department of Physics, Columbia University, 538 West 120th Street, New York, NY 10027-5255, USA\\

\end{affiliations}

\dedication{}

\keywords{Shimming, Dark Matter, Nuclear Magnetic Resonance, Bayesian Optimization}

\begin{abstract}
The sensitivity and accessible mass range of magnetic resonance searches for axionlike dark matter depends on the homogeneity of applied magnetic fields. Optimizing homogeneity through shimming requires exploring a large parameter space which can be prohibitively time consuming. We have automated the process of tuning the shim-coil currents by employing an algorithm based on Bayesian optimization. This method is especially suited for applications where the duration of a single optimization step prohibits exploring the parameter space extensively or when there is no prior information on the optimal operation point. Using the Cosmic Axion Spin Precession Experiment (CASPEr)-gradient low-field apparatus, we show that for our setup this method converges after approximately 30 iterations to a sub-10 parts-per-million field homogeneity which is desirable for our dark matter search.
\end{abstract}

\section{Introduction}

\subsection{Magnetic resonance searches for axions}\label{sec:axion}

Astronomical observations point at an abundance of dark matter (DM) in the universe which interacts gravitationally, but surprisingly, could not yet be explained within the Standard Model (SM) of particle physics. One class of particles beyond the SM that could explain this puzzling phenomenon are axions. Originally, the axion was proposed in 1977 by Peccei and Quinn as a solution to the strong $CP$ problem of quantum chromodynamics~\cite{PecceiQuinn, PhysRevLett.40.223, PhysRevLett.40.279}. In contrast to the weak interaction, the strong interaction conserves the discrete symmetries of a particle state under the operators $C$ (charge conjugation), $P$ (parity reversal), and $T$ (time reversal). In particular, this means that the combination $CP$ is also an unbroken symmetry. However, quantum chromodynamics (QCD) predicts that the strong force has a part that does violate CP symmetry in the form of vacuum field configurations, also called instantons, as it was found by Gerardus 't Hooft in 1976 \cite{PhysRevD.14.3432}. This CP-violating part was parametrized as an additional, non-perturbative term in the expression for the QCD's Action, or Lagrange density, featuring a vacuum angle $\theta$ as an open parameter. This was done in order to solve the so-called $U_{A}(1)$ problem of the strong interaction, however, it created a new puzzle: The strong force evidently not violating CP would correspond to a very small or nonexistent vacuum angle $\theta$, but QCD does not restrict this parameter, in theory there is an equal reason for it to be anywhere from $0$ to $2\pi$~\cite{Peccei}. This dilemma was subsequently dubbed the strong CP problem~\cite{MANNEL2007170}.\\
A direct consequence of the strong force violating CP would be that the neutron would posses a nonzero electric dipole moment of $d_n\approx 10^{-16}$~e cm~\cite{Dar:2000tn}. However, experimental results constrain the value of the quantity to $|d_n| < 2.9 \cdot 10^{-26}$~e cm~\cite{Baker_2006}.\\
As an approach to solving the enigma, R.Peccei and H.Quinn postulated the existence of a new global chiral symmetry $U(1)_{PQ}$~\cite{Peccei2008}. This Peccei-Quinn symmetry is spontaneously broken, leading to the existence of a new Nambu-Goldstone boson via the Goldstone theorem, which would have very low mass and very weak coupling to any other sector of the SM. As such, $\theta$ is essentially promoted to a field, with the associated boson filling the role of the CP-violating parameter $\theta$ and the low coupling of this particle naturally corresponding to a small CP-violation term in the strong interaction~\cite{PecceiQuinn}. Since then, various extensions have been introduced, some of which do not solve the $CP$ problem but could still be DM and are thus referred to as axionlike particles (ALPs).
If a significant fraction of the DM with a local density of $\rho_\mathrm{DM} \approx 0.4$\,GeV/cm$^3$ consists of ALPs with an unknown mass of $m_a$, their number density has to be so high that much of their collective behavior can be described as classical waves,

\begin{equation}\label{eq:wavyaxions}
    a(\Vec{r},t) = a_0 \cos{(\omega_a t - \Vec{k}\cdot\Vec{r}+\phi)}\,.
\end{equation}

These oscillate at the ALP Compton frequency $\omega_a= m_ac^2 / \hbar$, where $c$ is the speed of light in vacuum and $\hbar$ the reduced Planck constant. Furthermore, assuming ALPs with uniform mass, $a_0 \propto \sqrt{\rho_\mathrm{DM}}/m_a$ is the field's amplitude, $\Vec{k}=m_a\Vec{v}/\hbar$ the wave vector, and $\phi$ its phase. The coherence time of the field, determined by the velocity spread of the virialized DM, is given by $\tau_a \approx 10^6/\omega_a$.\\
\medskip ALPs are predicted to couple to photons, gluons and fermions in various ways~\cite{PhysRevD.88.035023}. An interaction between ALPs and fermions can be described by the Hamiltonian 

\begin{equation}\label{eq:hamil1}
    H_a = \hbar g_{aNN} \nabla  a\cdot\hat{\sigma}_N\,,
\end{equation}

where $g_{aNN}$ describes the gradient coupling strength and $\hat{\sigma}_N$ is a nuclear spin operator. This is analogous to the Zeeman interaction where a magnetic field acts on spin, in the nuclear spin case with a coupling strength determined by the gyromagnetic ratio $\gamma$. Therefore, the ALP field gradient $\nabla a$ can effectively be treated as a pseudo-magnetic field $\Vec{B}_a$ when acting on nuclear spin: $H_a = \hbar\gamma \Vec{B}_a \cdot \hat{\sigma}_N$. This can result in, for example, detectable perturbations of spin precession.\\
\medskip With tools such as optically pumped magnetometers and nuclear magnetic resonance (NMR) spectroscopy, several collaborations have probed the interaction between the ALP field and spins~\cite{GNOME, PhysRevLett.122.191302, PhysRevX.4.021030, Bloch_2022, karanth2023search, Ablikim_2023, PhysRevX.13.011050}. In this paper, we focus on the Cosmic Axion Spin Precession Experiments~\cite{kimball2018overview} sensitive to the gradient coupling (CASPEr-gradient). Here, a spin-polarized sample placed in a leading field $B_0$ can acquire a measurable transverse magnetization when the Larmor frequency $\omega_L = \gamma B_0$ is equal to $\omega_a$. An alternative approach relies on modulations of the Larmor frequency induced by the ALP field, resulting in sidebands in the NMR spectrum~\cite{Garcon_2017, Garcon_2019}. The affiliated CASPEr-electric setup also relies on NMR techniques, but instead probes the ALP-gluon coupling~\cite{aybasSearchAxionlikeDark2021}.

\subsection{Field homogeneity and the sensitivity to axionlike particles}

As is generally the case in NMR experiments, the sensitivity of CASPEr-gradient depends on, among other factors, the number of spins in the sample, the degree of spin polarization, as well as on the transverse relaxation time. In most search schemes, it is advantageous to increase the latter to be at least as large as the ALP-field coherence time so that the sensitivity scales as $T^{1/2}$ with measurement time $T$~\cite{PhysRevX.4.021030, Gramolin_2022, zhang2023frequencyscanning}. When applying the sideband technique \cite{Garcon_2017, Garcon_2019, PhysRevD.97.055006}, long relaxation times are also advantageous because in that case the lowest accessible ALP mass depends on how well the sideband can be resolved from the carrier.\\\medskip
The total transverse relaxation time, denoted by $T_2^*$ and defined as the time for the transverse magnetization to decay to 1/$e$ of the initial value, is determined by several effects. One is due to the fact that, in practice, the magnetic field across the sample is not homogeneous. Therefore, spins at different locations precess at slightly different Larmor frequencies resulting in dephasing parametrized by the relaxation time $T_\nabla$. Reducing the field inhomogeneity is typically done by applying suitable correction fields in a process called active shimming, which is the focus of this paper. Another common source of relaxation is due to the chemical environment of the sample which leads to, for example, spin-spin interactions causing dephasing. The relaxation time due to these effects is commonly denoted as $T_2$~\cite{PhysRev.70.460}. In some cases, relaxation time $T_d$ due to diffusion also has to be considered. Assuming that all the relaxation processes are purely exponential, one finds that the total transverse relaxation time is determined by

\begin{equation}\label{eq:T2components}
    \frac{1}{T_2^*} = \frac{1}{T_2} + \frac{1}{T_\nabla} + \frac{1}{T_d}\,.
\end{equation}

Therefore, at magnetic resonance searches for ALPs, it is often beneficial to improve the field homogeneity to minimize $1/T_\nabla$. 
Furthermore, ideally a sample with $T_2 > \tau_a$ is employed, although this sometimes conflicts with other desirable properties. For example, a solid-state sample with relatively short $T_2$ is used for CASPEr-electric, as the ferroelectric properties of the sample boost their sensitivity to ALP-interactions.\\
\medskip In this work, we focus on minimizing the resonance line width, dominated by $1/T_\nabla$, in the CASPEr-gradient experiment employing liquid samples. Ultimately, we aim for a line width similar to that of the ALP signal, which is expected to be a few part-per-million~\cite{zhang2023frequencyscanning}.

\section{The CASPEr-gradient-low-field apparatus}\label{sec:CASPEr}

\begin{figure}
    \centering
    {\includegraphics[width=.8\linewidth]{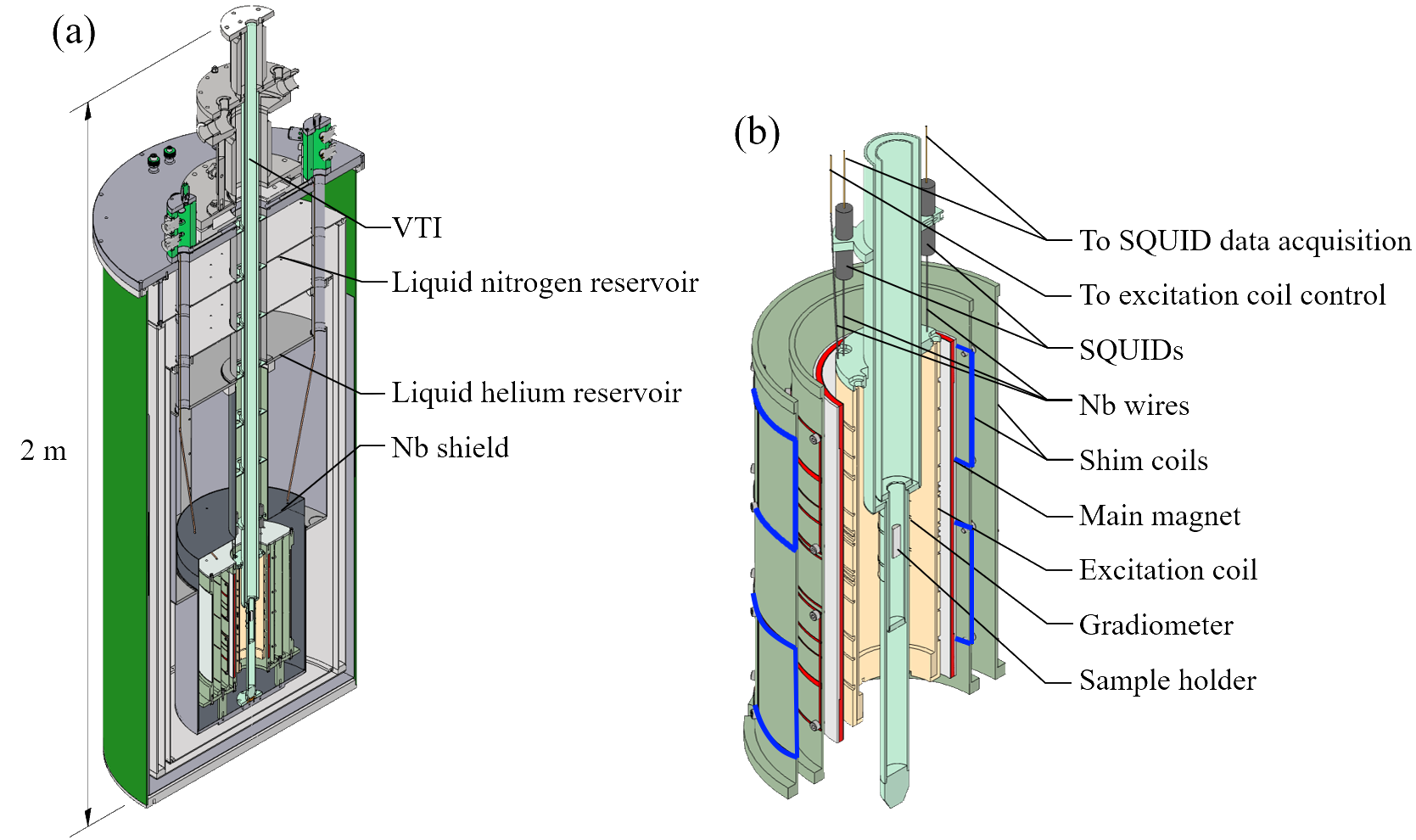}}
    \caption{Cross sections of the CASPEr-gradient LF apparatus. (a) View of the complete cryostat with the superconducting coil structure near the bottom of the liquid helium reservoir. The isolation vacuum of the variable temperature insert (VTI) allows for sample temperatures in a range of $160~\mathrm{K}$ to $300~\mathrm{K}$. (b) Detailed view of the coil structure and sample region. Red (blue) lines highlight the main (shim) coils. The SQUIDs, kept in niobium housings, are placed higher than shown so as to not be affected by the main magnet field.}
    \label{fig:casper2}
\end{figure}

The results presented in this work were obtained with the CASPEr-gradient low-field (LF) setup and will also be applied at the high field setup which is currently under construction. The LF setup consists of a liquid-helium cryostat containing a superconducting solenoid capable of generating a leading field of up to $B_0 \leq 0.1$~T. Figure~\ref{fig:casper2} shows a schematic drawing of the cryostat. The apparatus has a built-in superconducting shield against external magnetic fields and is placed in a dedicated Faraday room to reduce the effects of stray electromagnetic radiation on the detector electronics.

Measurements presented in this work are performed on approximately 1.2 mL of thermally polarized liquid methanol (CH$_3$OH) in a cylindrical volume. This is placed at the center of the leading field coil using a variable temperature insert (VTI) so that its temperature can be kept above the freezing point. For measuring the sample magnetization transverse to the leading field, a first-order gradiometer made of 50~$\mathrm{\mu}$m thick niobium wire is installed on the outside of the VTI.\\\medskip
The flux in the gradiometer is coupled to superconducting quantum interference devices (SQUIDs) operated in flux-locked-loop mode. Its signals are fed out of the Faraday room to a digital lock-in amplifier for processing and recording. For purposes such as shimming and diagnostics, a set of transverse coils tunable with a T-circuit is installed. Pulses are applied using a Magritek Kea2 NMR spectrometer. Our simulations show that, without shimming, the field distortions due to the superconducting wire of the gradiometer are at the level of 30 parts-per-million (ppm). 

\begin{table}
\centering
\begin{tabular}{c|c|c}
\hline
Shim coil & Field gradient direction & Order\\
\hline
 A & $x$ & 1 \\
 B & $z$ & 1 \\
 C & $y$ & 1 \\
 D & $z^2$ & 2 \\
 E & $xy$ & 2 \\
 F & $zx$ & 2 \\
 G & $zy$ & 2 \\
 H & $x^2-y^2$ & 2 \\
\hline
\end{tabular}
\caption{Shim coils with adjustable current input and the equations approximately describing the gradients of their fields~\cite{shimmanual}. The leading field $B_0$ is oriented along the $z$-direction. In this context, for example $x$ is a shorthand for the gradient component $\frac{\partial B_z}{\partial x}$. The order refers to the exponent in the $B_z$ field equation. Field maps of the generated fields can be found in, for example, the works by Romeo and Hoult~\cite{Romeo1984Mar} or Jayatilake et al.~\cite{FASTMAP}.}
\label{table:shimchannels}
\end{table}

\subsection{Magnet coils operation of the CASPEr-gradient setup}

\begin{figure}
    \centering
    {\includegraphics[width=0.7\linewidth]{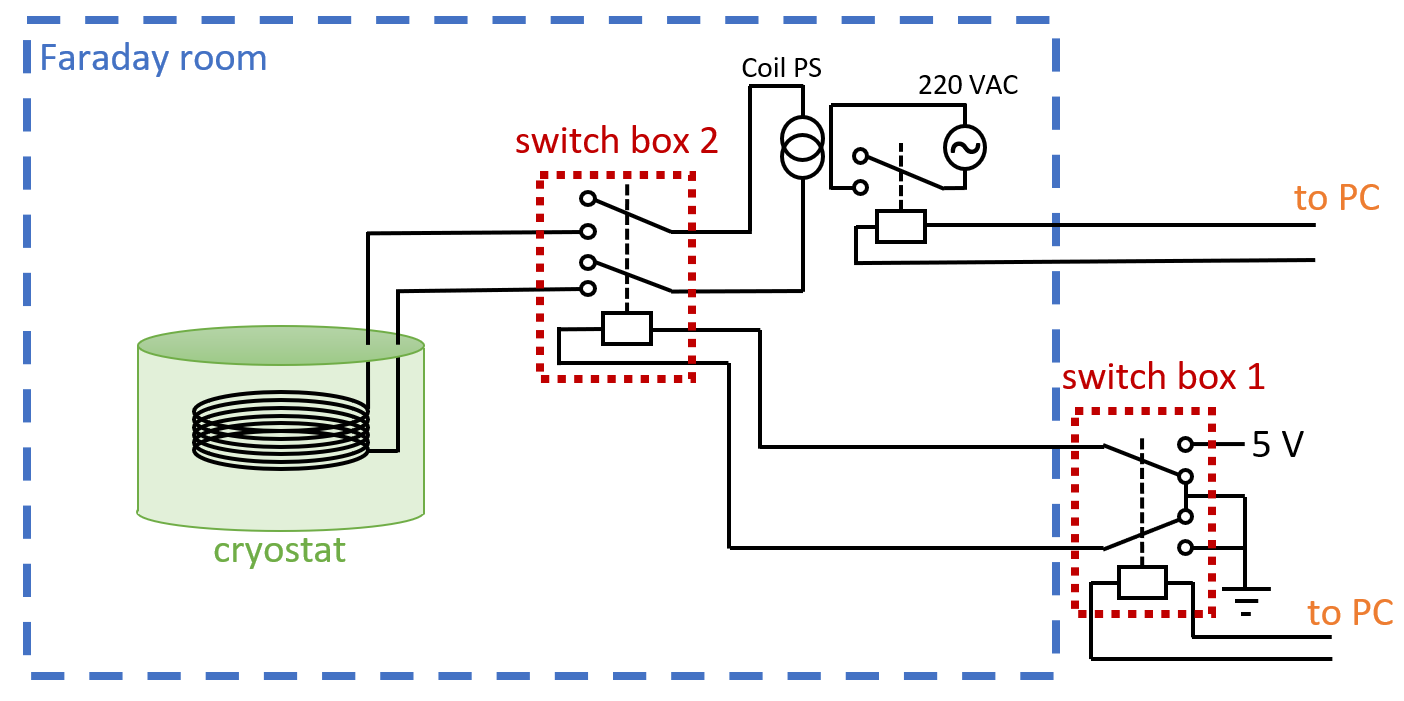}}
    \caption{Schematic of the remote switching electronics for the cryogenic coils. Only one coil, without cryogenic switch, is shown for clarity. In reality all coils and their cryogenic switches can be remotely disconnected using switch box 2. Switch box 1 uncouples switch box 2 from the PC in order to further reduce the noise introduced to the Faraday room. The $220~\mathrm{V}$ power for the coil current supplies is filtered before entering the Faraday room.}
    \label{fig:switchbox}
\end{figure}

CASPEr-gradient LF has eight shim coils\,\footnote{In high-field NMR, one is only interested in the gradients of $B_z$. There are three independent first-order gradients and five independent second-order gradients. The latter can be seen from the facts that the gradients do not depend on the order of coordinates over which differentiation is taken and that each of the field components obeys the Laplace equation, providing an additional relation among the second-order-gradient components.} which can be set in the range of approximately -2 to +2~A, refer to Table~\ref{table:shimchannels} and Figure~\ref{fig:casper2}. With these, a field homogeneity of at least 2~ppm over a spherical volume with diameter 8~mm can be achieved. However, this was only demonstrated before installation of the VTI with the gradiometer and excitation coils which introduce additional gradients. 
NMR measurements need to be performed with the coil power supply (PS) disconnected from the coil leads because even if it is switched off, the resultant noise picked up by the sensitive SQUID is too large. For this purpose, a set of remote controlled switch boxes was developed, see Figure~\ref{fig:switchbox}. These physically interrupt the connections between PS and coils as well as the coil heater switches by means of electromagnetic relays. Given the typical situation where the shim coils are in so-called persistent mode, changing the currents then follows the following procedure: 1) switch box 1 establishes a connection between PS and coils 2) the PS is switched on 3) the PS is ramped to the known currents in the coils 4) the heater switches are switched on 5) the PS ramps the coils to the new current 6) the heater switch is switched off so that the coils are back in persistent mode 7) the PS current is ramped to 0~A 8) the PS is switched off 9) the switch box physically disconnects the PS from the coils and heaters. This procedure takes approximately 30~s, mainly due to necessary settling time after operating the cryogenic heater switches.

\section{Methods to improve the field homogeneity}

Finding the set of shim coil currents that maximizes the homogeneity over the sample (known as shimming) is a common task in NMR experiments. It is performed, for example, when the sample geometry is modified or the leading field strength is changed. The latter is necessary during ALP searches at CASPEr-gradient in order to probe different masses. Typically, an iterative approach is applied where the shim current settings for the next step are informed by the results of an NMR measurement. Doing this manually is a tedious task requiring a significant time investment even of experts capable of deducing the geometry of the field inhomogeneity from the NMR line shape~\cite{SAM}. Alternatively, 3D gradient-echo pulse sequences can be applied to map the field and guide the shimming. Therefore, in many modern NMR spectrometers, some form of automatic shimming is implemented. This often relies on the simplex algorithm which is robust and well-suited for combinations of low-order and high-order shims but typically requires many steps~\cite{simplex}~\cite{9410624}. Recently, other alternatives have been investigated, such as a deep learning approach~\cite{BECKER2022107151}. In this work we use a Bayesian optimization algorithm with the goal of reliably obtaining good shim settings in a small number of iterations.\\\medskip
Our optimization goal was to reach signals with a fractional line width of 10~ppm or better. Given our setup we estimate this to be a sensible margin below which effects other than inhomogeneous broadening start to dominate the signal and sensitivity to dark matter.

\subsection{Bayesian optimization}\label{sec:bayes}

Bayesian optimization (BO) sequentially samples an unknown target function $f$ to construct and improve a surrogate of it. It is a derivative-free method without the need for initial assumptions about the functional shape of $f$. BO starts by placing a prior probability distribution over the function, which can be informed by an initial evaluation of $f$ at a single point. In our case, a Gaussian process prior is used. Following Bayes' rule, each optimization step consists of evaluating $f$ at a query point and then updating the prior with the data gathered through this sampling to calculate the posterior probability distribution. The next most efficient query point is computed based on the posterior, and the posterior acts as the prior in the next iteration.\\

\medskip BO is typically favored when optimising unknown, computationally intensive or very noisy functions~\cite{BOreview, 8957442}. It has applications in a wide range of problems in machine learning, networks and deep learning~\cite{brochu2010tutorial}, and many fields of natural science, such as particle physics~\cite{Ilten_2017}, chemistry and material design~\cite{BOchemistry}. By judiciously selecting query points, BO often needs fewer steps than derivative-based optimization methods~\cite{10.5555/3221315.3221581}~\cite{PracticalBayesopt}. In particular, compared to the widely used simplex algorithm, BO possesses properties promising to be more favorable for a fast shimming procedure: While the simplex polgygon moves through the parameter space of shim currents along a continuous path with a well-defined step size, BO can at any step choose probing points from anywhere within parameter space. In cases where the optimum lies far from the starting point, BO is thus able to potentially reach best shim settings faster than the simplex method. Furthermore, simplex algorithms often reduce their step size upon an improvement in the quality parameter, and are therefore prone to getting stuck at local maxima. BO on the other hand can adjust the degree of exploration of parameter space it undertakes, and thereby strike a balance between an accurate approximation of the target function and quickly finding extrema.\\
\medskip In our case, we define the target function by the following properties: $f_{shim}(I_A,...,I_H)$ takes eight variables $I_{A \cdots H}$ from within the parameter space of acceptable shim currents as input and outputs a single real-valued number. It is considered a black-box function for which an analytical expression is not known. Each query to $f_{shim}$ equals an NMR measurement. Its return value should be a measure for the quality of our NMR signal. When decreasing field inhomogeneity and thus increasing spin relaxation time $T_2^*$, we expect higher signal power and a more narrow signal line width $\Gamma$ in the frequency spectrum~\cite{YZMasterThesis}. Therefore we choose the signal peak amplitude, reconstructed from a fit to the spectrum, as the target.\\

\medskip In contrast to other optimization strategies, BO calculates the most useful query point at each step by predicting $f(x)$ in advance, before actually evaluating it. The prediction is based on the values that the posterior, constructed from all previously gathered data takes. All potential query points $x$ are associated with a utility function $\alpha(x)$ (also called acquisition function or infill sampling criterion). This function acts as an estimate for the predicted improvement associated with each query point, and is calculated after every sampling $t$. The maximum of the $t^{th}$ utility function will be selected as the $(t+1)^{th}$ sampling point:

\begin{equation}
    x_{t+1} = \mathrm{argmax}_{x,t}(\alpha(x,t))\,.
\end{equation}

In the following, some frequently used utility criteria which were tested in this work are introduced, and the quantities are related to the shimming scenario at hand. 
An uncomplicated approach is the Upper Confidence Bound (UCB) utility function,

\begin{equation}
    \alpha_{UCB}(x,\kappa) = \mu(x) + \sigma(x)\kappa\,.
\end{equation}

It returns for any combination of shim currents $x$ a simple weighted sum of expectation $\mu$ and uncertainty $\sigma$ of the posterior at that point, that is, the estimate for the quality factor returned if an NMR measurement is performed with settings $x$. We have introduced a positive factor $\kappa$, which can be seen a measure for the confidence level of the prediction.
It balances the degree of exploration the algorithm will undertake: Increasing $\kappa$ raises the likelihood of the algorithm choosing more uncertain points, possibly leading to more iterations but avoiding getting stuck on local maxima.\medskip

For introducing the next utility criteria, it is useful to define improvement $I$ between an optimization step $t$ and the following one, $t+1$, as:

\begin{equation}\label{eq:improvement}
   \mathcal{I}(x,t) = \max(f(x_{t+1}) - f(x_t^*), 0)\,.
\end{equation}

Here, $f(x_t^*)$ refers to the maximal value of $f(x)$ found at any step up to $t$.
The Probability of Improvement (POI) criterion is used to find for a point $x_{t+1}$ the probability of $f(x_{t+1})$ being larger than the current maximum $f(x_t^*)$, without considering the magnitude of the improvement. Our Gaussian process posterior suggests that at any point $x$, the unknown value of $f(x)$ is sampled from a normal distribution $\mathcal{N}$ with mean $\mu(x)$ and variance $\sigma^2(x)$,

\begin{equation}
    f(x) \sim \mathcal{N} (\mu(x), \sigma^2(x))\,.
\end{equation}

We now use the reparameterization~\cite{AcqiFunc} $f(x) = \mu(x) + \sigma(x)z$, where $z$ is distributed as $\mathcal{N}(0,1)$ (with a probability density function of $\varphi(z)=\frac{1}{\sqrt{2\pi}}\exp{(-z^2/2)}$).
Then the probability of a positive improvement when probing point $x$, $P(\mathcal{I}(x)>0) \Longleftrightarrow P(f(x)>f(x^*))$, as predicted by the posterior, can be understood as the area enclosed by the Gaussian curve associated with $x$ cut off at $z > z_0$, $z_0 = \frac{f(x^*)-\mu(x)}{\sigma(x)}$. Injecting again the exploration parameter $\kappa$, it can be calculated using the cumulative distribution function (or upper-tail probability) $\Phi(z)$:

\begin{gather}\label{eq:acquisi_POI}
    \alpha_{POI}(x,\kappa) = P(f(x) \geq f(x^*) -\kappa)) \\
    = 1 - \Phi(z_0) = \Phi(-z_0) = \Phi \left( \frac{\mu_(x)-f(x^*)+\kappa}{\sigma(x)} \right)\,.
\end{gather}

The maximum of the Expected Improvement (EI) distribution $\alpha_{EI}(x)$ corresponds to the point with highest expected improvement over the current maximum. That is, it returns the shim setting predicted to result in the highest signal amplitude according to the current model of our black-box function. The EI function is evaluated as the expectation value of $\mathcal{I}(x)$:

\begin{equation}
    \alpha_{EI}(x,z) = \int_{-\infty}^{\infty} \mathcal{I}(x,z) \varphi(z) dz\,.
\end{equation}

Since $\mathcal{I}(x,z)=0$ for $z<z_0$ with $z_0=\frac{f(x^*)-\mu}{\sigma}$, the expression reduces to:

\begin{equation}
    \alpha_{EI}(x,z) 
    = \int_{z_0}^{\infty} (\mu + \sigma z - f(x^*)) \varphi(z) dz\,.
\end{equation}

This integral can be evaluated as:

\begin{equation}
    \alpha_{EI}(x,\kappa) 
    = (\mu - f(x^*)-\kappa) \Phi\left( \frac{\mu - f(x^*)-\kappa}{\sigma} \right) + \sigma\varphi\left( \frac{\mu - f(x^*)-\kappa}{\sigma} \right)\,.
\end{equation}

Utility functions can prioritize exploration (favor points of high uncertainty, typically far from previous training points) or exploitation (favor points of high expectation for a local maximum, close to the most successful training points) at various ratios. The advantage of BO lies in the fact that computing and maximising the utility function is in most cases drastically faster and more efficient than evaluating $f$ itself. Bayes' rule is used each time a query to the function has been performed, to update the covariance $\sigma(x)$.

\subsection{Shimming routine}\label{sec:autoshim}

The results presented in this work were obtained using the following shimming routine:

\begin{enumerate}
    \item Set suitable initial conditions.
    \item Apply one or more 90-degree pulses and record the resulting SQUID signal.
    \item Calculate the frequency spectrum and determine the peak amplitude through a fit.
    \item Update the surrogate function with the new data point.
    \item Maximize the utility function to determine settings where the predicted improvement is largest.
    \item Ramp to the new shim current settings.
    \item Repeat from step 2 on until the desired signal quality is reached or until improvement falls below a set threshold.
\end{enumerate}

All shimming sequences start from a shim setting that results in a visible, but relatively small and broad NMR peak. These settings are found by an experimenter by trial and error: Usually, acceptable settings are searched for one shim coil at a time in up to 5 tries per coil. Step 1 therefore requires some work from the user, which can be minimal once reasonable shim settings are known from previous shimming sequences. Sometimes an early step of automated shimming reduces the signal to below the noise threshold again. Ideally, the algorithm finds the signal back by reverting to shim settings close to those that were previously successful, but this depends on the value of the exploration parameter $\kappa$. In some cases, the algorithm misidentifies a noise peak as signal and will unsuccessfully attempt to enhance it. Therefore, the end result of an automated shimming run should still be assessed by the user to make sure the signal is valid.\\
\medskip To test a given configuration of shim fields, we apply a 90-degree pulse with the excitation coils and record the transverse magnetization using the SQUID. From this free induction decay (FID) the averaged power spectral density (PSD) is calculated. In our case, the goal of the BO procedure is to optimize the maximum power of the NMR signal. The total power in the spectrum, which is proportional to the relaxation time, is simply given by the integral of the squared FID, but in case of a low signal-to-noise ratio (SNR) there is the risk of optimizing noise power instead of that of the signal. Instead, an attempt to fit the PSD with a Lorentzian on top of constant noise background is made. The extracted Lorentzian amplitude $S=A / \Gamma$, where $A$ refers to the enclosed area and $\Gamma \propto \frac{1}{T_2^*}$ to the line width, is used as a measure for the power of the signal. This method has proven to be the most robust one for our setup where the SNR is low. The disadvantage is that it relies on assumptions regarding the line shape, e.g., when multiple peaks are to be expected due to chemical shifts the fitting function has to be adapted. For this reason, in NMR setups sometimes deuterated solvents are admixed to the sample in order to generate well-known frequency-shifted proton lines that can be used as reference. In our case, we do not aim to obtain sub-ppm resolution so that fitting a single Lorentzian is sufficient. It should be noted that while our black-box function assumes a dependence of the resulting line shape only on the applied shim currents, in reality signal quality depends also on whether the conditions of the sample, for example temperature, remain constant over the measurement duration.\\
\medskip All steps but the first were automated. The algorithm's code uses an existing Python implementation of BO~\cite{bayesopt}. Its optimization logic consists of a sequence of the methods \emph{suggest}, \emph{probe} and \emph{register}: \emph{suggest} proposes a parameter set to probe, which is the utility function's maximum. It uses the scikit-learn~\cite{scikit-learn} GaussianProcessRegressor class to construct the surrogate function. 
\emph{probe} evaluates the function at the query point passed by \emph{suggest}. \emph{register} adds the points probed and target value obtained by \emph{probe} to a dictionary storing all accrued information about the target function. From this, the surrogate is updated. We run the algorithm for at least 30 steps, which takes approximately half an hour. The algorithm supports more manipulation of the optimization sequence such as constraining parameter space at each step, but we did not use this feature for this work.

\section{Results}\label{sec:results}

We present our exploration of the utility functions introduced in section~\ref{sec:bayes}, followed by an investigation of how the balancing parameter $\kappa$ influences the effectiveness with which the optimal shim currents are found. All test runs shown in Figures~\ref{fig:ACQcomp} and~\ref{fig:UCBcomp} started from reasonable shim settings, which reliably produced signals with a fractional NMR line width of approximately 20~ppm. This reflects the reality where reasonable but non-optimal shim settings are known from experience with the particular system, or by scaling the shim currents proportionally to the leading field. 

\subsection{Test of various acquisition functions}

\begin{figure}
    \centering
    {\includegraphics[width=\linewidth]{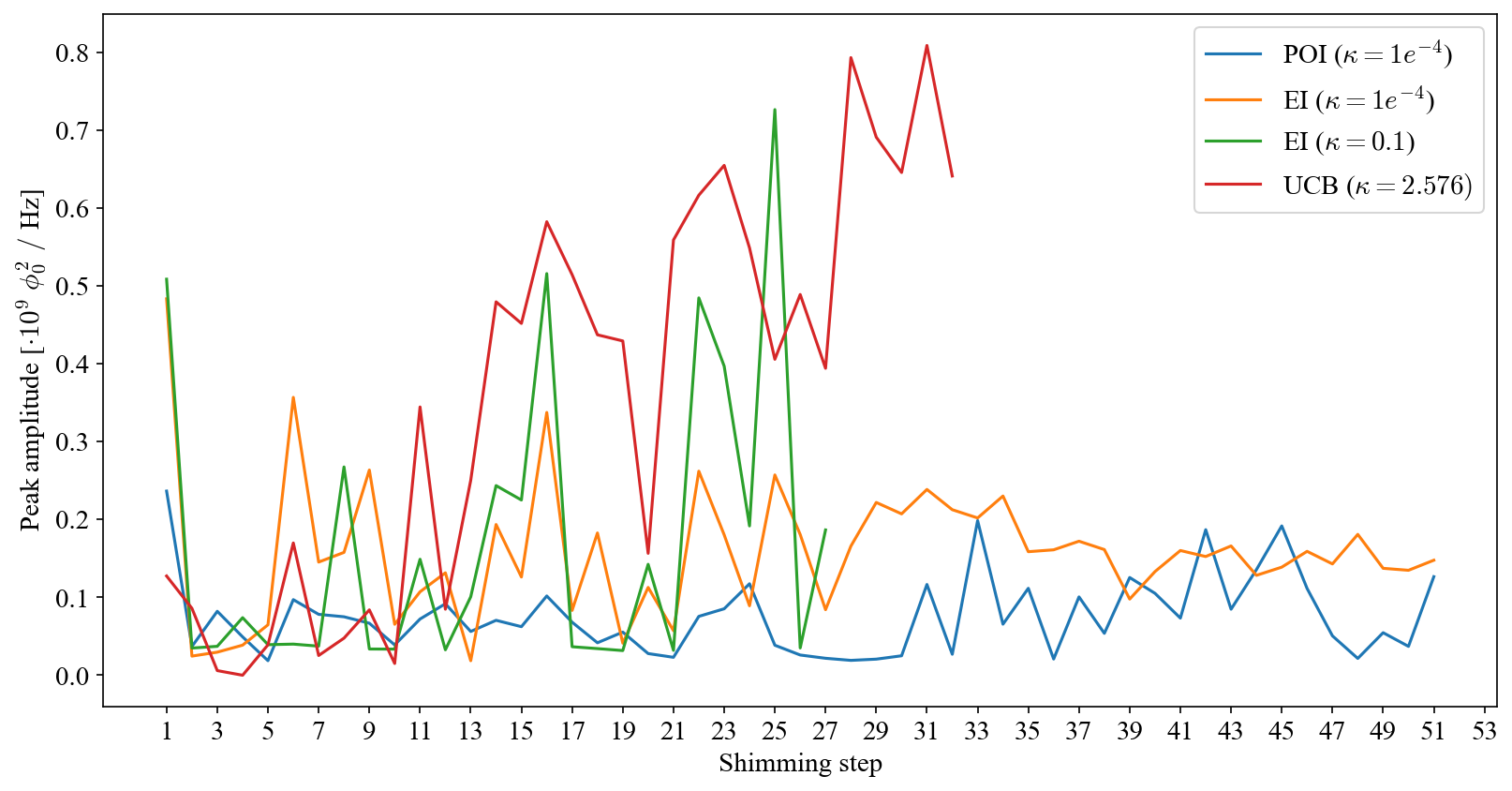}}
    \caption{Fitted peak amplitude evolution over $\approx$~30 shimming steps, for different acquisition functions. In the two cases where the algorithm was not able to find shim currents better than the initial settings, the run was extended to 50 iterations.}
    \label{fig:ACQcomp}
\end{figure}

The acquisition criteria UCB, EI and POI were tested with the exploration parameters set to the default values suggested by the authors of the BO implementation, namely $\kappa_{UCB}=2.576$, $\kappa_{EI}=\kappa_{POI}=1\cdot10^{-4}$. In addition a run with EI acquisition and a higher exploration factor $\kappa=0.1$ was performed. The number of optimization steps were chosen so that each test lasted a maximum of one hour, allowing for many tests to be done. Furthermore, this is in line with our goal of employing an algorithm that finds suitable settings fast so that more time can be spend on our dark matter search. The resulting changes in fitted peak amplitude are shown in Figure~\ref{fig:ACQcomp}. Even though all runs started from identical initial shim currents,  the initial peak amplitudes vary because not all relevant parameters could be kept constant. For example, because the sample is taken out during breaks, its position and temperature is not sufficiently reproducible. For this and other reasons, a more advanced temperature control system at CASPEr-gradient LF is under development.\\
\medskip Only with the UCB acquisition function and EI acquisition with exploration parameter 0.1 an increase of peak amplitude was achieved. With the latter run, the amplitude fluctuates more strongly than with EI and $\kappa=1\cdot10^{-4}$, reflecting the higher degree of exploration undertaken. The test with POI resulted in the loss of the signal which was not recovered. Even though this acquisition function selects the points which are most likely to yield an improvement according to the surrogate model, in reality not each step in this run came with an increase in amplitude. This shows that with the given amount of data, the constructed model was not yet a reliable approximation of our target function $f_{shim}$. In this work, we do not aim to fully explore all reasonable acquisition functions and their exploration parameters. Instead of trying to obtain better results with the POI function by changing $\kappa$, we chose to further investigate the UCB acquisition function due to its promising initial performance and its straightforward exploration to exploitation trade-off.

\subsection{Exploration versus exploitation setting of the UCB criterion}

\begin{figure}
    \centering
    {\includegraphics[width=\linewidth]{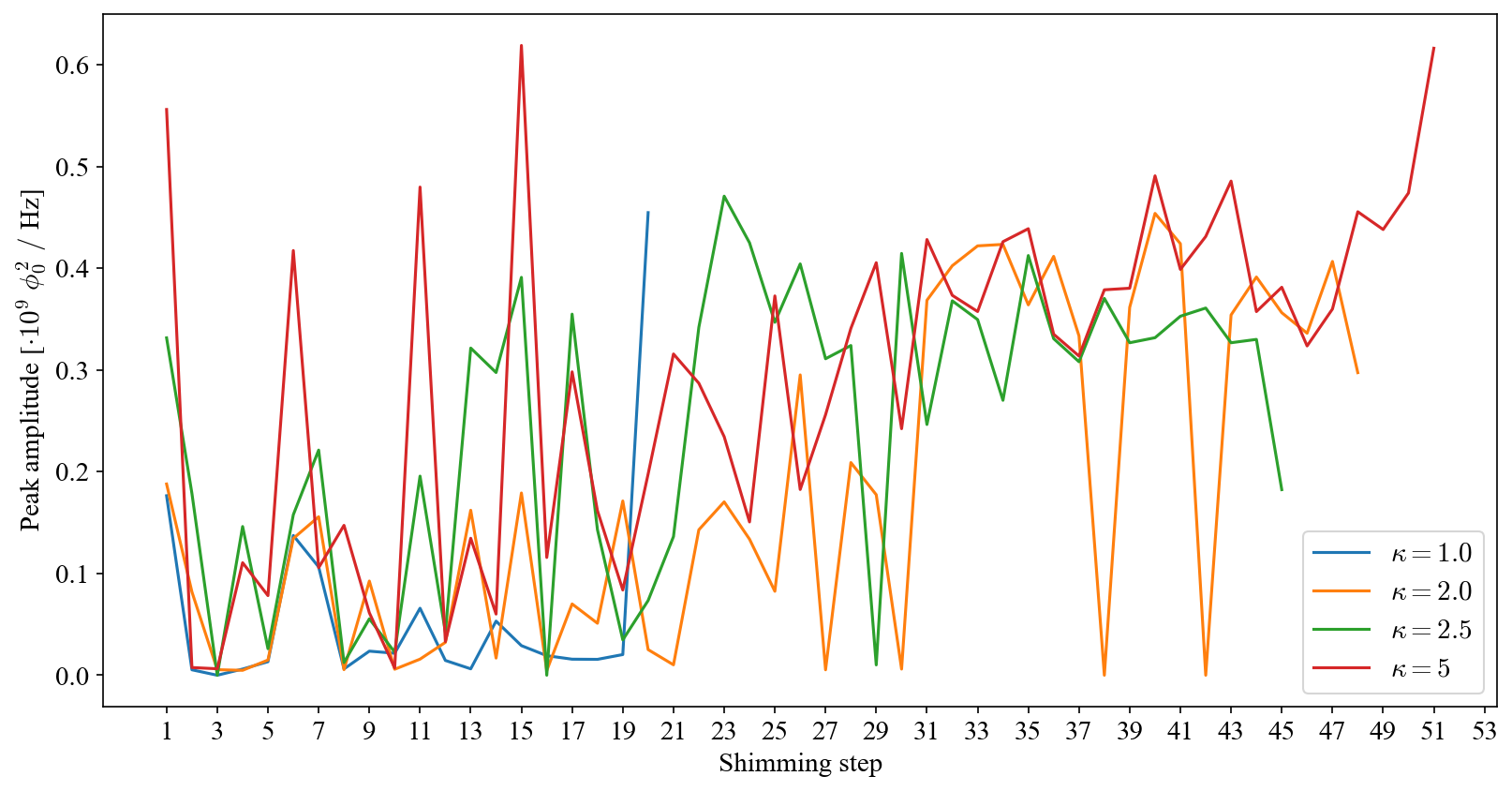}}
    \caption{Peak amplitude evolution for UCB acquisition with different exploration factors $\kappa$. At some points, the Lorentzian fit of the NMR peak did not yield reliable results so that the amplitude was defined as zero. Some runs were stopped after less than $50$ steps due to time constraints for the testing.}
    \label{fig:UCBcomp}
\end{figure}

We performed five test runs each over 50 steps with values $\kappa\in\{1.0;1.5;2.0;2.5;5.0\}$, all starting from the same initial currents. These values were chosen to test a wide range of degrees of uncertainty on the posterior. Namely, as seen from equation~(5), they corresponds to 68, 87, 95.4, 98.76, and 99.99994~\% uncertainty. The maximal peak amplitudes achieved in these runs are plotted in  Figure~\ref{fig:UCBcomp} shows the peak amplitude evolution over time for four of the test runs. Another run with $\kappa=10$ was started from zero shim currents as an attempt to test a very high degree of exploration. Due to showing little continuous improvement over 20 iterations, this run was cancelled. The run with $\kappa=1.0$ crashed after 20~steps due to technical problems. It yielded the highest total improvement of 158~\%, however the average improvement per step for this setting is low. As the largest improvement occured only at the last step, the algorithm may have been stuck at a local maximum for most of the time. In addition, this amplitude gain could not be reproduced and instead in a later run with the same settings, only a gain of 40~\%, comparable to $\kappa=2.0$, was achieved. $\kappa=5$ resulted in the highest overall amplitude, with a low improvement. This is in part due to the already very high initial amplitude. For the majority of shimming runs listed in Table~\ref{table:shimrecords} we have used a setting of either $\kappa=2.0$, or $\kappa=2.576$ (corresponding to a $99\%$ confidence interval). From our limited empirical evidence, the algorithm seems to perform most reliably around the $\kappa=2.5$ mark. Overall it is fairly robust with regards to the exploration setting, but will generally require more steps for higher $\kappa$ values.


\subsection{Results of BO-guided shimming with UCB acquisition}

\begin{figure}
    \centering
    \subfigure[]{\includegraphics[width=.9\linewidth]{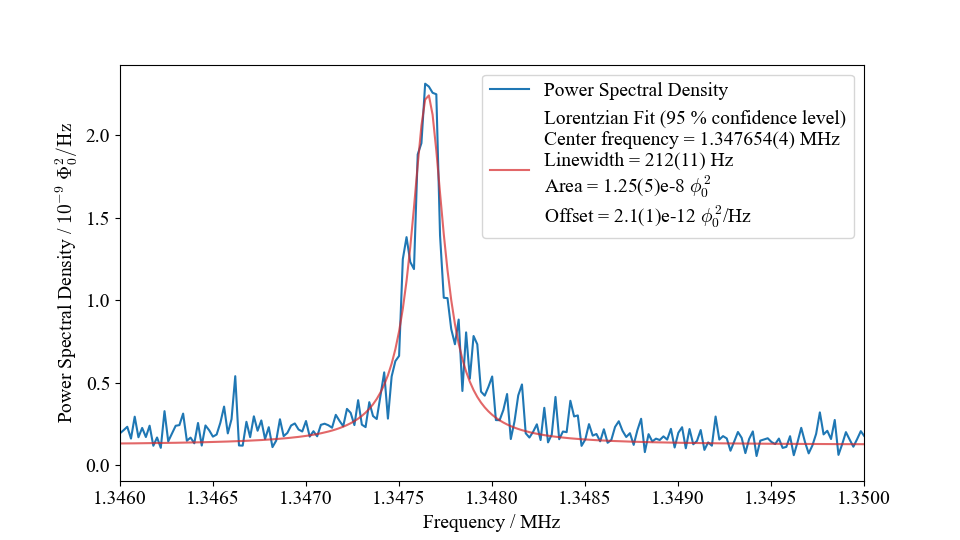}}
    \subfigure[]{\includegraphics[width=.9\linewidth]{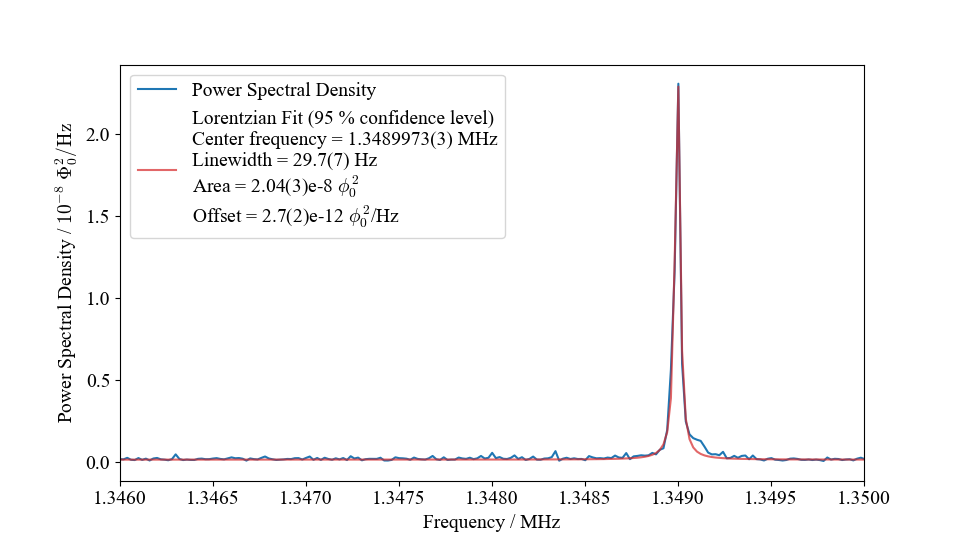}}
    \caption{(a) PSD spectrum over frequencies recorded before a typical shim run, zoomed to a frequency band between $1.346$~MHz and $1.350$~MHz. The parameters of the Lorentzian fit to the peak are given, uncertainties are denoted in brackets. (b) The spectrum after $15$ steps of BO-guided shimming.}
    \label{fig:beforeafter}
\end{figure}

\begin{figure}
    \centering
    {\includegraphics[width=.9\linewidth]{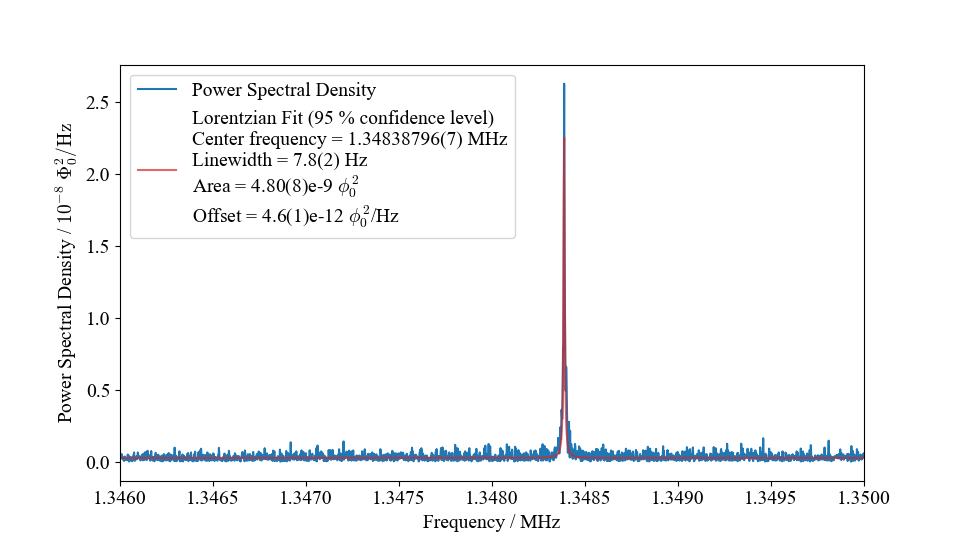}}
    \caption{Overall best shimming result achieved at CASPEr-gradient so far with BO-guided shimming. These settings were the result of running our algorithm five subsequent times over a total of 123 iterations, each run starting from previous best settings. Due to the Lorentzian fit not accurately representing the signal peak, a percentile filter was instead used to determine the peak amplitude as $2.62(43)\cdot 10^{-8}\phi^2_0/Hz$. Because this spectrum was taken on a different day than those shown in Figure~\ref{fig:beforeafter}, the experimental conditions were not exactly identical and the noise and signal amplitude should not be directly compared.}
    \label{fig:bestresult}
\end{figure}

\begin{figure}
    \centering
    {\includegraphics[width=0.65\linewidth]{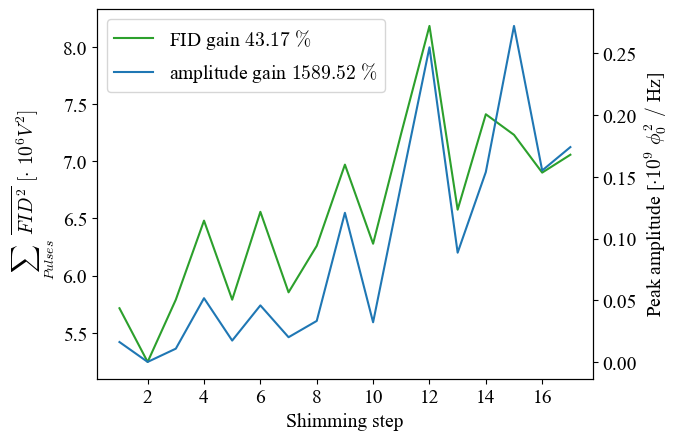}}
    \caption{Evolution of peak amplitude as a function of shimming steps for the run presented in~\ref{fig:beforeafter}.}
    \label{fig:qualityevol}
\end{figure}

\begin{figure}
    \centering
    {\includegraphics[width=\linewidth]{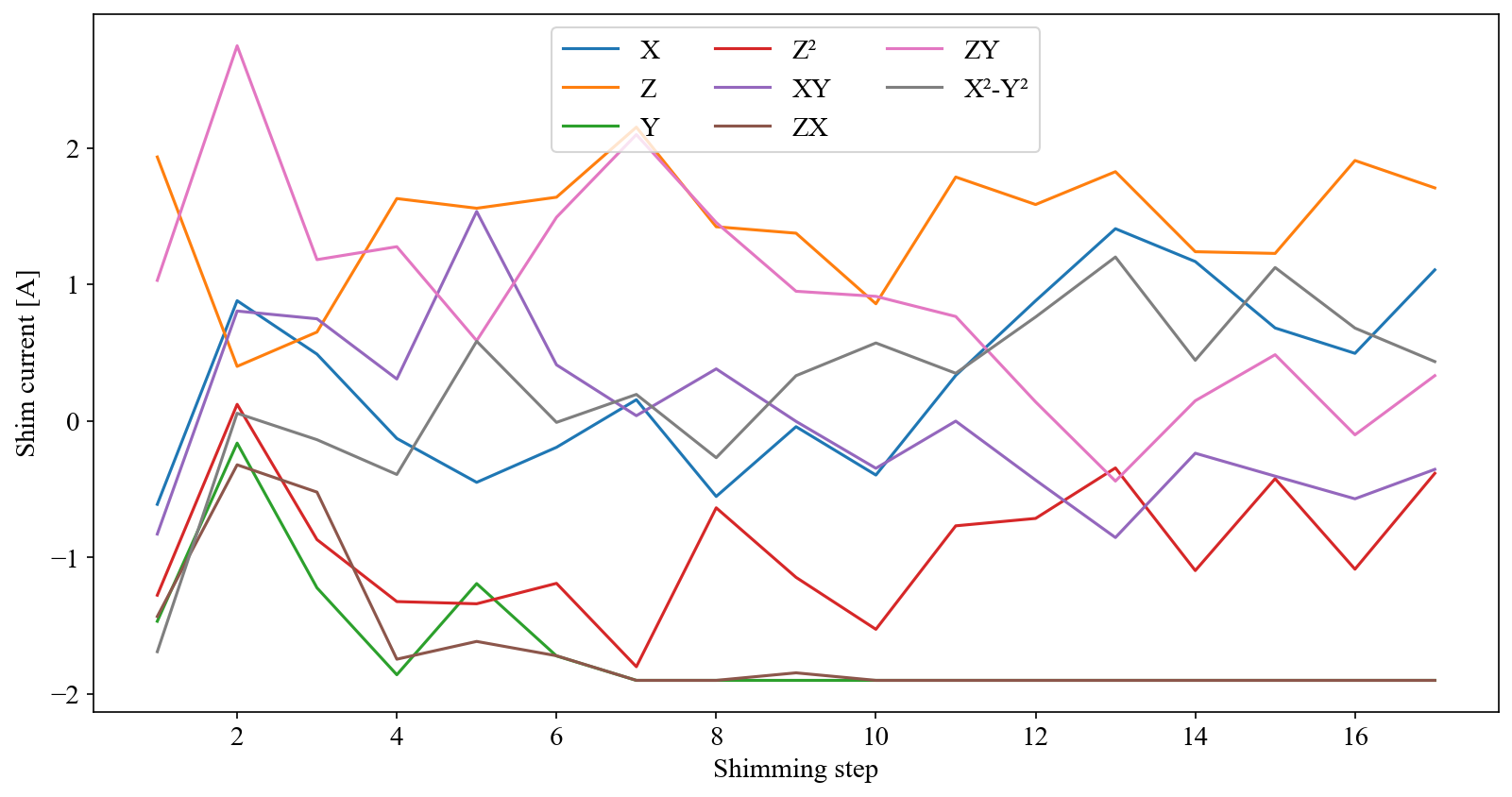}}
    \caption{Evolution of shim currents~\ref{table:shimchannels} as a function of shimming steps for the run presented in~\ref{fig:beforeafter}.}
    \label{fig:currentsevol}
\end{figure}

A total of 87 BO-guided shim sequences performed at CASPEr-gradient LF were considered for this work, most of them using a UCB acquisition function with $\kappa = 2.5$. Both the quality of results and efficiency of the shimming process could be greatly increased compared to user-guided shimming. On average, BO shimming sequences lasted for 29 iterations. In one run, the optimal settings were found after an average of only 15 iterations. The smallest NMR signal line width achieved, $7.8 (2)$~Hz, is showcased in Figure~\ref{fig:bestresult}. Here and throughout the text, the numbers in brackets indicate the 1 standard deviation uncertainties on the least significant digits. The total best and average values, as well as the relative gain per single optimization step and more statistics can be found in Table~\ref{table:shimrecords}.\\
\medskip Figure~\ref{fig:beforeafter} depicts the result of a typical shimming sequence with UCB acquisition, $\kappa=2.5$ and 17 optimization steps. The spectrum at the start of the run and the one after 15 iterations, where best shim currents had been found is shown. In this particular run the position of the NMR peak has been shifted from 1.347\,654(4)~MHz to 1.348\,997(2)~MHz, due to the shim fields components in $z$-direction changing the sample's overall Larmor frequency. This remains well within the region of good impedance matching of the setup. Peak line width could be reduced by a factor of approximately 0.14, or 86~\%, from 213~Hz to 30~Hz. Signal amplitude could be increased by almost a factor of 12, resulting in a drastically better SNR. For this run the time evolution of both the square sum over recorded FIDs and the signal amplitude reconstructed from a Lorentzian fit to the NMR signal peak are plotted in Figure~\ref{fig:qualityevol}. They exhibit similarities in their shape, confirming theoretical expectation as both are proportional to the power contained within the recorded signal. In total, the run caused an increase in both FID and peak amplitude which are our alternative measures for signal quality. The zig-zag behavior of shim currents over time, see Figure~\ref{fig:currentsevol}, showcases the algorithm's exploration. The allowed parameter range for the algorithm to explore was constrained to $\{-1.8,1.8\}~\mathrm{A}$ to prevent shim coils from quenching.\\
\medskip At the corresponding leading field level, a total of four more shimming cycles were performed following the one discussed above, each starting at the best currents found in the previous one. Optimal settings in these runs were found at steps $\{28;22;16;42\}$. The third of these was done with EI acquisition, $\kappa=0.1$ and doubled as the test run for this function shown in Figure~\ref{fig:ACQcomp}. The others were run with UCB acquisition, $\kappa=2.0$. After these runs the signal line width had been reduced to 7.8(2)~Hz, exceeding our shimming goal of sub-10 ppm line width. The final spectrum is plotted in Figure~\ref{fig:bestresult}.

\begin{table}
\centering
\begin{tabular}{|c|c|}
\hline
Record & Value \\
\hline
Number of shimming runs & $87$ \\
Average shimming steps per run & $29$ \\
Average steps until best settings & $15$ \\
Smallest peak line width [Hz] & $7.8(2)$ \\
Average final peak line width [Hz] & $160.76(59)$\\
Highest line width reduction in a run [\%] & $89(2)$ \\
Average line width reduction per run [\%] & $2.50(60)$ \\
Highest peak amplitude [$10^{-8}\mathrm{\phi}_0^2/$Hz] & $3.90(10)$ \\
Highest amplitude gain in a run [\%] & $17292(415)$ \\
Average amplitude gain in a run [\%] & $583(14)$ \\
Highest amplitude gain in a step [\%] & $720(17)$ \\
Average amplitude gain in a step [\%] & $24.80(60)$ \\
\hline
\end{tabular}
\caption{Shimming achievements at CASPEr-gradient using the BO method described in this article. The numbers in brackets indicate the 1 standard deviation uncertainties on the least significant digits.}
\label{table:shimrecords}
\end{table}

\FloatBarrier

\section{Conclusion}

The sensitivity of magnetic resonance searches for ALPs depends on the relaxation time of the sample spins, which is partially determined by the homogeneity of the applied magnetic field. We developed methods to find optimal shim coil currents in an automated way in order to reduce leading field inhomogeneities at CASPEr-gradient LF. Due to hardware limitations, our priority was to perform the shimming in as few steps as possible. This is important in ALP searches where the leading field is scanned and consequently the shim currents have to be regularly adjusted. Bayesian optimization proved to be robust and efficient, and using an upper confidence bound acquisition function with exploration factor near $\kappa=2.5$ was found to produce reliable results for our setup. Possibly, further optimization of the parameters of the algorithm can yield even better performance, but our goal of reducing the NMR line width to below 10~ppm was achieved within drastically shorter time compared to our previous strategy of manual searches for good shim settings. The algorithm is generally applicable to NMR spectrometers with shim coils where it can be a useful alternative to common methods that either need significant input from experienced users, are prone to getting stuck at local maxima, or need many iterations to converge.

\section*{Acknowledgements}

This work was supported in part by the Cluster of Excellence ``Precision Physics, Fundamental Interactions, and Structure of Matter'' (PRISMA+ EXC 2118/1) funded by the German Research Foundation (DFG) within the German Excellence Strategy (Project ID 39083149) and COST Action COSMIC WISPers CA21106, supported by COST (European Cooperation in Science and Technology).

\printbibliography

\end{document}